# Mechanical and Systems Biology of Cancer

Fabian Spill[1+], Chris Bakal[2], Michael Mak[3+*]

[1] School of Mathematics, University of Birmingham, Birmingham, B15 2TT, UK
[2] Division of Cancer Biology, The Institute of Cancer Research, London, SW3 6JB, UK
[3] Department of Biomedical Engineering, Yale University, New Haven, USA
+ equal contribution
* correspondence: Michael.Mak@Yale.edu

**Manuscript Information**
   Number of Figures: 3

# Abstract

Mechanics and biochemical signaling are both often deregulated in cancer, leading to cancer cell phenotypes that exhibit increased invasiveness, proliferation, and survival. The dynamics and interactions of cytoskeletal components control basic mechanical properties, such as cell tension, stiffness, and engagement with the extracellular environment, which can lead to extracellular matrix remodeling. Intracellular mechanics can alter signaling and transcription factors, impacting cell decision making. Additionally, signaling from soluble and mechanical factors in the extracellular environment, such as substrate stiffness and ligand density, can modulate cytoskeletal dynamics. Computational models closely integrated with experimental support, incorporating cancer-specific parameters, can provide quantitative assessments and serve as predictive tools toward dissecting the feedback between signaling and mechanics and across multiple scales and domains in tumor progression.

## Introduction

The mechanical microenvironment in cancer is vastly altered compared to healthy tissue. Typically, the extracellular matrix (ECM) is stiffened in the tumor microenvironment (1-3), but individual cancer cells may actually be softer (4). There is a bimodal distribution of nanomechanical stiffness across advanced cancer tissues (5). Moreover, more complex mechanical and geometric characteristics, including the fibrous matrix structure, porosity, or viscoelastic parameters may be changed in tumors (6, 7). Similarly, solid and fluid stresses are greatly altered in cancers (8). It is well known that cancers exhibit increased fluid pressures, in part due to remodeling of the vasculature and lymphatics (9).

The altered ECM stiffness and geometry of the tumor microenvironment are sensed by tumor cells via mechanosensing structures, which can activate intracellular signaling pathways that drive behaviors such as unrestrained proliferation, increased survival, tissue invasion, stemness, and drug resistance (10-12). While cancer has been traditionally considered a genetic disease, alterations in ECM stiffness and geometry can force normal cells to adopt phenotypes characteristic of transformed and/or metastatic cells in the absence of any genetic change (13, 14). Theoretical work suggests that environmental cues, coupled with various possible oncogenic alterations (e.g. overexpression of c-Src (15)), can drive cancer progression (16, 17). Cancer progression can be promoted by genetic changes that alter how cells respond to ECM stiffness and geometry and that enable cancer cells to remodel their environment in ways that promote disease.

To open new therapeutic avenues that seek to manipulate the response of cancer cells to their environment as a way to treat cancer, predictive mathematical models are required to describe how cell fate decisions are due to interactions between tumor cells and their ECM and how these interactions differ between normal and cancer cells. The problem is inherently multiscale in nature and involves diverse components such as biochemical reactions, cell-matrix and cell-cell interactions, and tissue-level alterations. The field of mechanotransduction has long embraced modelling tools in order to describe how cells respond to mechanical and geometric cues, and these models serve as key starting points for more complex descriptions of how cancer cells interact with their ECM. For example, models have been developed that provide insights into diverse aspects of mechanobiology including: force-dependent molecular bonds (18-21), spatiotemporal organization of intracellular molecules (22-24), impact of cell shape (25-29), and the dynamics of the cytoskeleton (30-32). Here we review some of these models and supporting experimental findings with a look toward the future. We first review recent work on cytoskeletal interactions that modulate intracellular mechanics and the propagation of cytoskeletal forces inside and outside the cell. Next we focus on the cell-matrix adhesion complexes that act as key signal transducers and mechanosensors. Finally, we review key signaling networks implicated in mechanotransduction.

## Generation and propagation of intracellular forces

The active actin cytoskeleton provides basic structure and force generation capabilities. The key components include actin filaments, actin crosslinking proteins (ACPs) such as alpha-actinin and filamin, and myosin II motors that generate contractility. Inside the cell, a large network of these components undergoes dynamic and stochastic interactions, spontaneously resulting in pattern formation – including the actin cortex at the cell periphery, thick contractile bundles of actin (stress fibers), and cell polarity (leading and trailing edges). Local interactions and kinetics can control overall, global functionality of the cytoskeletal network. In particular, actin turnover rates can modulate cytoskeletal network tension, and the interplay between actin turnover, actin crosslinking, and myosin II walking activity can regulate the morphological state of the network, from homogeneous morphologies to local clusters (Fig. 1a) (30). Computational simulations can isolate individual features and determine their roles in cytoskeletal network behavior. For example, altering actin nucleation rates can modulate the stress fluctuation magnitudes in the cytoskeleton, a phenotype observed in intracellular microrheology experiments that modulate epidermal growth factor (EGF) signaling (known to influence actin nucleation) in breast cancer cells (33). Additionally, spatial and temporal profiles are important in regulating cell behavior. These can be precisely tuned in computational models. For example, cell geometry and dimensionality influence the anisotropy and amplitude of intracellular stress fluctuations (34). While overall cell tensions have an intuitive role of enabling cells to apply forces onto their substrate (e.g. the ECM) and migrate, intracellular stress fluctuations can facilitate the redistribution of organelles and molecular components inside the crowded cytoplasmic space (35). Furthermore, malignant tumor cells appear to exhibit larger intracellular displacement and stress fluctuations compared to benign counterparts, as shown by experiments measuring intracellular stiffness and force fluctuations (35). Cytoskeletal mechanics and fluctuations are the result of the interactions between many cytoskeletal components, each undergoing dynamic processes (turnover, walking, binding, unbinding, etc.). Computational network models of the cytoskeleton, based on physical principles (reaction kinetics, mechanics) and incorporating realistic, experimentally tangible features, can help dissect the local, molecular-level contributions to experimentally observable mechanical cellular phenotypes. High resolution experimental techniques, e.g. super resolution imaging or atomic force microscopy, can help guide the development and validation of models of fine and distinct cytoskeletal features (36). Furthermore, models coupling cytoskeletal forces to critical intracellular and extracellular features, particularly the nucleus and the ECM, can start to elucidate a more holistic picture of cell behavior.

Cytoskeletal forces can be transmitted to the cell nucleus via the LINC (Linker of Nucleoskeleton and Cytoskeleton) complex (37). Substrate stiffness modulates cytoskeletal tension and thus nuclear stress and shape, which interestingly also modulates the expression levels of a key nucleoskeletal protein lamin A, nuclear stiffness, and stem cell differentiation (38). The mechanical properties of the nucleus can also influence nuclear shape and dynamics during cell deformation and invasion through confined spaces (e.g. ECM pores or endothelial junctions). Large nuclear deformations can lead to rupture and DNA damage, as observed in experimental studies of cancer cells invading through highly confined constrictions (39, 40). Computational models coupling cellular forces to the nucleus can generate quantitative details of nuclear deformation and mechanical remodeling during physiological processes and draw insights toward differences in nuclear behavior due to biochemical or structural alterations. For example, experiments show that lamin A/C deficiency leads to more plastic remodeling of the nucleus after larger strains, which can be captured in a continuum model of the nucleus featuring a hyperelastic shell and a poroelasto-plastic core (Fig. 1b) (41). Furthermore, the role of different types of lamins (A and B) in regulating nuclear shape and geometry can be explored in continuum models through incorporating heterogeneous material profiles. In particular, a preferred mesh size difference between lamin A and lamin B appears to explain nuclear blebbing tendencies (42).

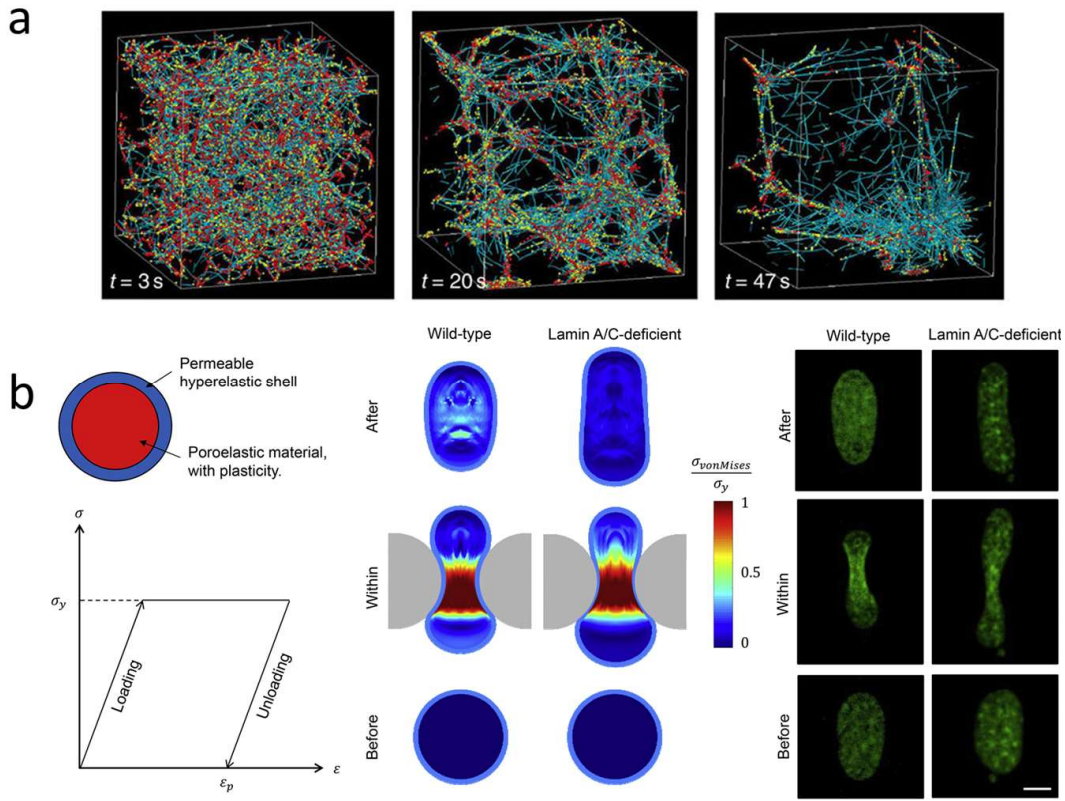

Figure 1: **Computational models of cell mechanics.** a) Brownian dynamics simulations of the active actin cytoskeleton demonstrate cytoskeletal network evolution, a process dependent on the interplay between actin turnover, network crosslinking, and myosin activity. The simulation domain is 3x3x3μm$^3$ with periodic boundary conditions. Actin filaments are teal, myosin II motors are red, and actin crosslinking proteins are yellow. Adapted from (30). b) A finite elements model of the nucleus predicts stress profiles and plastic remodeling after deformation through a confined barrier, e.g. endothelial junction. The model is composed of a permeable hyperelastic shell surrounding a poroelastic-plastic core. The color bar indicates relative stress levels. The green fluorescence experimental images (right) show nucleus morphologies at different stages during the deformation process. Adapted from (41).

In many types of solid tumors, cancer cells are embedded in a dense fibrillar matrix. Cytoskeletal forces are transmitted into the ECM via cell-matrix adhesions, which can lead to ECM remodeling and propagate mechanical signals to surrounding cells (43). Stiffer substrates tend to promote increased cell traction forces and lead to a more invasive phenotype (44, 45). Relaxation of tension in the substrate in laser ablation experiments (46, 47) tends to revert cell invasiveness. Moreover, ECM networks exhibit nonlinear strain stiffening (48), suggesting potential mechanical feedback mechanisms. These phenomena have been demonstrated through a number of experimental studies. Complementarily, computational models can provide quantitative, mechanistic insights toward underlying driving factors of invasive behavior in 3D ECMs – particularly to a level of detail that may be unfeasible for experiments to achieve or parse out. Computationally intensive models can capture a high degree of local details observed in high resolution experiments of cell-ECM interactions. In a recent study, a model capturing an entire cell with dynamic protrusions inside a surrounding ECM showed that dynamic filopodia can act as rigidity sensors that facilitate durotaxis in HUVECs (Fig. 2a) (49). While stiffness sensing (and many other cell behaviors) is a phenomenon exhibited by normal and cancer cells, cancer-related parameters can be tuned in generalizable models to explore disease phenotypes. In particular, the above model showed that the number and length of filopodia can modulate invasive behavior, supporting prior studies that showed that deregulation in filopodia-related functions and pathways are implicated in cancer progression and metastasis (50). In another model that incorporates dynamic local forces and force-sensitive ECM fiber-fiber crosslinks, it is demonstrated that the coupling of mechanical forces and fiber-fiber biochemical kinetics can result in ECM densification near the cell boundary, consistent with experiments in tumor and endothelial cells (51). Furthermore, the fibrillar nature of the ECM and asymmetric contractility of elongated cells can lead to long range anisotropic strain profiles in the environment due to fiber realignment (Fig. 2b) (52-54), which can also generate spatial profiles of stiffness (48).

Tumors often grow as large multicellular masses in which cell-cell and tumor-ECM interactions as well as environmental properties can dictate cancer progression. Computational models of collective tumor invasion and evolution have been developed that aim to capture patterns observed in clinical data (e.g. from histology or clinical databases). Mathematical, biophysical relationships (e.g. in the form of partial differential equations for continuum features and/or rules and probabilities for discrete features) can be used to govern the behavior and spatiotemporal profiles of tumor content, consisting of a mixture of components (cells, ECM, fluid, concentration profiles of nutrients, chemokines, and drugs, etc.) (55). For instance, the role of adhesions on invasion or growth was investigated in the models (56-58) and the role of angiogenesis in tumor growth was modeled in (59). Some models are able to capture overall tumor geometries seen in clinical data (60) as well as provide insights toward complex factors influencing drug response (55). An important next step is the integration of multi-physics tumor models with more realistic biophysical features in the tumor microenvironment and associated signal transduction networks and signaling mechanisms. We refer the reader to (61, 62) for recent reviews specifically focusing on multicellularity and tumor modelling.

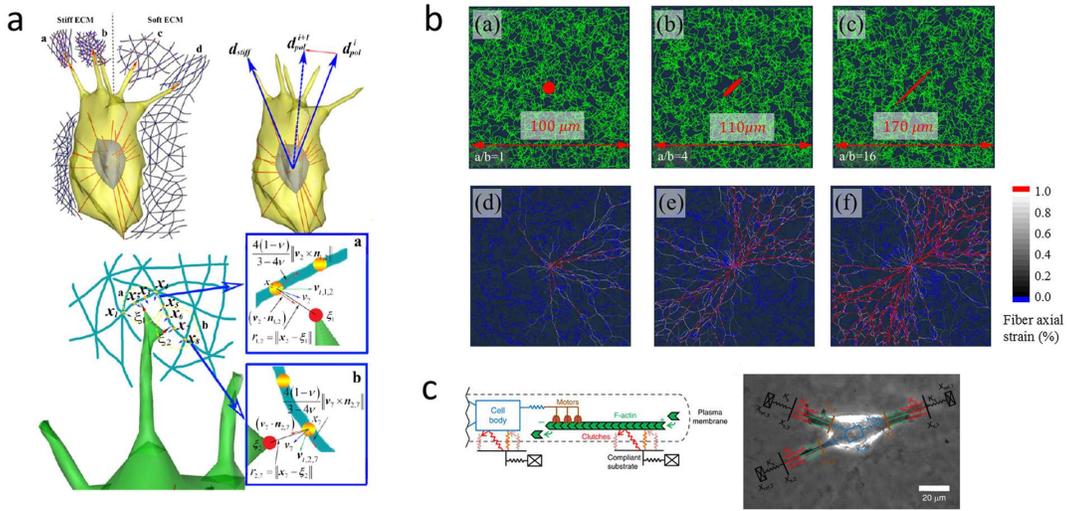

Figure 2: **Models of cell-matrix interactions.** a) A 3D cell model with stress fibers, a nucleus, and filopodia captures mechanical cell-matrix interactions and indicates a potential role for filopodia in stiffness sensing. Filopodia can protrude, adhere to ECM fibers, and contract, pulling fibers and sensing their stiffness. The cell body will tend to polarize more toward stiffer regions. Adapted from (49). b) An ECM fiber model shows network strain distribution (bottom) as a function of cell contraction anisotropy (top). More spindle-like cells, which tend to contract more along one axis, can generate farther reaching anisotropic strain fields in the fibrillar ECM network. Adapted from (54). c) Multiple motor-clutch components are used to model a cell migrating on a deformable substrate, predicting an optimal substrate stiffness to maximize cell migration speed. Clutches bind and unbind in a force-dependent manner from the substrate, and molecular motors retract F-actin which is connected to the clutches, thus pulling the substrate. Adapted from (79).

The physical environment surrounding solid tumor cells is dynamic and heterogeneous, influenced by the presence of cancer and stromal cells (63). Precise spatial and temporal physical profiles (of stiffness, architecture, ligand density, etc.) of this environment can be explored through mechano-chemical models that interface active cells with a responsive, physiologically mimicking ECM. These profiles in turn can act as signals that cells can sense through complex mechanisms mediated by adhesion complexes. Elucidating detailed signal transduction effects then requires models that couple mechanics with biochemical signaling networks.

## Focal adhesion dynamics, mechanosensing, and signaling

Focal adhesions (FA) are multifunctional organelles that serve as primary points of sensing of ECM stiffness and geometry by cells (64). FAs are much more than passive receptors, but rather are dynamical systems comprised of complex interactions between the ECM, the cytoskeleton, and signal transduction machinery across multiple spatiotemporal scales. The emergent behavior of these systems underpins the generation, transmission, and coordination of diverse forces, changes in cell shape, and cell fate determination, including the acquisition of malignant and therapy-resistant phenotypes (65).

At the heart of FAs are membrane bound integrins (Fig. 3b), (66). As cells interact with their ECM, for example via protrusions generated by actin polymerization at the leading edge, the binding of individual integrin molecules to the ECM initiates – integrin clustering; the activation of FAK; the

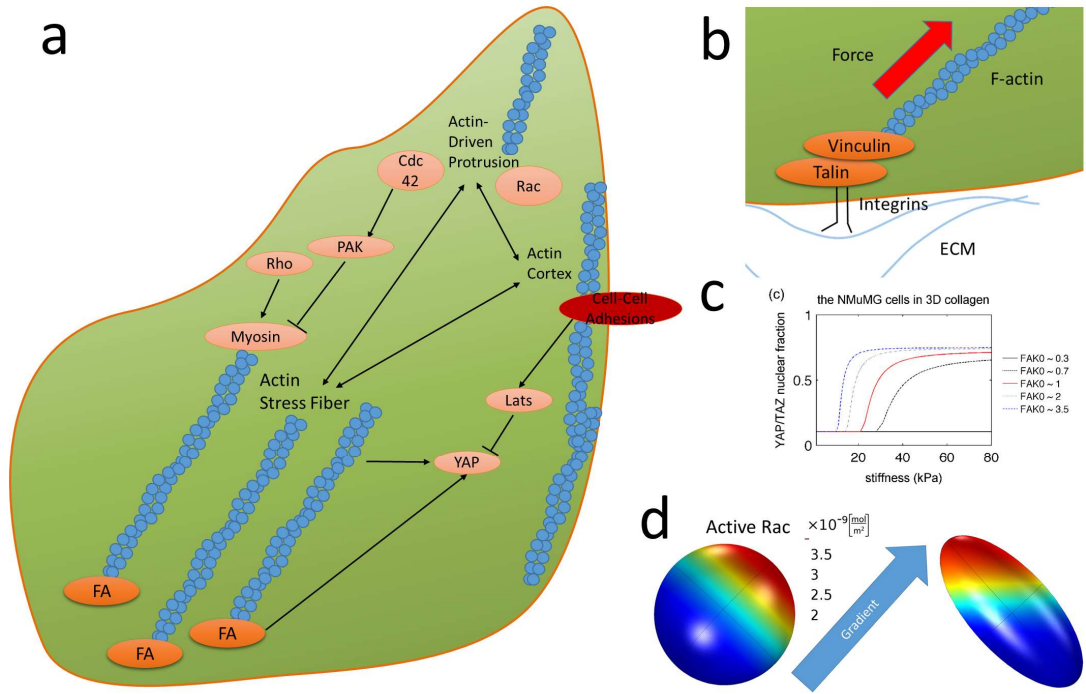

Figure 3: **Mechanosensing and mechano-regulating pathways.** a) Different cytoskeletal compartments are regulated by different pathways, compete with each other, and affect different downstream effectors. Cdc42 and Rac mediate different actin-driven protrusions, whereas Rho activates actin stress fibers. Different cytoskeletal compartments (e.g. cortex, stress fibers, protrusions) may compete for the global actin pool. Therefore, strengthening one compartment may weaken others. Moreover, crosstalk of regulators, e.g. through PAKs, may directly lead to competition of these compartments. Focal adhesions (FAs) are complex structures linking the cytoskeleton and the surrounding extracellular matrix. Some of the complexity are shown in (b). b) Integrins provide a direct attachment from inside the cell to the outside matrix. Inside, they link to the cytoskeleton through proteins including talin and vinculin that are force-sensitive. Consequently, forces may affect chemical reactions and thus adhesion assembly and disassembly. c) Focal adhesion mechanosensing will then lead to downstream effects on phenotypes, mediated, for instance, through YAP/TAZ. Focal adhesion kinase (FAK) is shown to shift the stiffness response of YAP/TAZ. Adapted from (110). d) Cell shape may also affect intracellular signaling directly. A model shows that the round cell adapts cytoskeletal regulators (here Rac) in the direction of the external stimulus (e.g. chemotactic gradient), whereas the ellipsoidal cell polarizes in a direction in between its longest axis and the stimulus. Adapted from (28).

subsequent recruitment of the proteins such as paxillin, talin, and vinculin; and the formation of nascent FAs. Talin and vinculin also bind actin in branched networks that are actively flowing over adhesions, which results in the transmission of force to the ECM, allowing cells to probe the stiffness of the ECM and generate traction (67-69). However, because engagement of actin by FAs prevents polymerized actin from generating further protrusions, in order to maintain protrusiveness, FAs transiently disengage from the actin network, allowing polymerized actin to slide by FAs and continue to push on the leading edge (70). This ability to engage and disengage actin networks by FAs has been termed the actin-FA "clutch". Increased clustering, further recruitment of molecules which couple FAs to actin, and post-translational events can lead to the maturation of FAs at sites distal from the leading edge. As adhesions mature, the nature of the actin organization at the FAs also differs, as stress fibers predominate on more mature adhesions; which can propagate relatively large forces throughout the cell body, leading to large morphological changes. During the formation of both nascent and mature

FAs, both mechanical and biochemical processes occur which will ultimately trigger their turnover, and thus from a systems-perspective, FA dynamics involve both extensive positive feedback loops (i.e. initial activating events such as integrin clustering become amplified), and negative feedback loops (i.e. FA formation leads to force generation and upregulates signals that will ultimately induce FA turnover).

Iteration between modelling and experimentation has yielded deep understanding of FA dynamics. The FA clutch concept was first described theoretically (71), which was followed by experimental observations of actin flow and FAs in living cells, (70, 72, 73). These works then led to the development of a stochastic model of the clutch (74) that predicts the existence of a regime of fast retrograde flow with low traction forces and one with slow retrograde flow and high traction forces, with extracellular stiffness acting as the switch between the regimes. More recent models have been developed which incorporate dynamic interactions between the individual FA components talin and vinculin (75) and which incorporate information regarding the spatial distribution of individual ECM-integrin-clutches (76). Furthermore, mechanochemical models that include feedback between adhesion assembly and substrate rigidity demonstrate different possible regimes of focal adhesion evolution, from nascent, unstable adhesions to stable adhesions with a steady-state size (77), and the interplay of ECM stiffness, remodeling, and ligand density can influence focal adhesion size and growth (78). A recent model with multiple motor-clutch modules in a cell-scaled geometry demonstrated the number of motor-clutch constructs in the cell influences the optimal substrate stiffness for maximal cell migration, supported by experiments in glioma cells (Fig. 2c) (79). Taken together, these and other studies are providing the first types of multi-scale models that explain how cellular phenotypes emerge from the dynamical interplay between FAs and the ECM.

Importantly, signaling events play both a key role in regulating short term FA dynamics (seconds-minutes) and in regulating the organization of the actin cytoskeleton around FAs. A simplified schematic linking important signaling pathways with cytoskeletal features is illustrated in Fig. 3. In particular, signaling via Rho GTPases such as Rac1, Cdc42, and RhoA play essential roles regulating the relationship between FAs and actin (Fig. 3a), (80). These roles have been particularly well studied in the context of migrating cells. For instance, chemotactic signals may lead to the activation of Rho GTPases that are known to regulate the cytoskeleton in different ways (81, 82). Moreover, forces may also activate Rho, through ROCK and myosin, leading to further forces pulling on focal adhesions and thus their reinforcement (83). Mechanical signals such as stiffness gradients consequently activate mechanoregulatory pathways, resulting, for instance, in durotaxis, i.e. the migration along stiffness gradients (84). Mathematical models were developed linking adhesion dynamics and durotaxis and showed that cell velocity depends on stiffness in a non-monotonic way, with a maximum at an intermediate stiffness (85). Collectively migrating cells may also durotax due to cells deforming the substrate more in the low stiffness regions (86), in line with experimental observations (87). Other models focus on the intricate details of the interplay of stress fibers and focal adhesion dynamics. For instance, the interplay of stress fibers and adhesion bonds was investigated (88, 89), and it was found that cyclic stretch may induce cell reorientation through reduction of the catch bond lifetimes of focal adhesions (90).

Cell shape, which emerges from the spatiotemporal dynamics of FA generation and turnover, can also impact both FA and cytoskeletal dynamics and signaling events (Fig. 3d), (25, 91). Cell shape can be considered a geometric cue and, like the spatial organization of the ECM-integrin complexes themselves (92), is important to consider in order to understand how cells respond to mechanical cues in the environment. Because cell morphological dynamics are dysregulated in many cancer types, mechanosensing by FAs may be affected, which can be explored through coupling cell geometries with cytoskeletal and FA kinetics in models. Cell-scaled mechanochemical models that include spatial profiles of tensional components inside the cell, particularly stress fibers, and substrate adhesions can reproduce cell shapes and stress distributions comparable to experimental studies (89, 93, 94).

## Signaling downstream of mechanical stimuli and feedback on mechanics

FAs serve as a platform for the assembly of large signaling complexes which regulate a host of downstream processes, particularly transcription. Transcriptional changes can influence cell-wide behaviors over much longer terms (hours-days); or even have permanent consequences (differentiation). These longer-term changes in cell state can modulate short term FA dynamics.

A classic example of how short term FA dynamics, in response to ECM properties (mechanical cues) and geometric cues (ECM organization and cell shape), drives long term changes in cell fate is how entry into the cell cycle by adherent cells is dependent on cell spreading. Observations by Dulbecco and Folkman provided the first evidence of the link between mechanosensing and proliferation (95-97). Following work demonstrated that spreading upregulated ERK and RhoA activity which upregulates the transcription of pro-proliferative factors CyclinD1 and downregulates pro-quiescence factors such as p21 and p27 (98, 99). Although the role of adherence in driving proliferation is not fully understood in cancer cells, recent studies, demonstrating that FAK is a key mediator of resistance to inhibitors of ERK activators, strongly hint FA mediated activation of ERK is an important driver of tumorigenesis (100, 101).

More recent work has shown that the YAP and TAZ transcriptional co-activators are also effectors of signaling complexes formed at FAs (75, 91, 102). Regulation of transcriptional events by YAP/TAZ is involved in a broad number of cellular behaviors that are essential drivers of tumorigenesis and metastasis, including proliferation, maintenance of stemness, and migration (103-105). Intriguingly, the mechanisms by which YAP/TAZ is activated appear to be highly dependent on the type of adhesion. At very early adhesions, YAP/TAZ is activated largely by focal adhesion kinase (FAK) through PI3K and/or mTOR (106-108). At nascent adhesions, FAK remains important for YAP/TAZ activation, and this activation appears to rely on signaling via the ARHFGEF7/beta-Pix Rho GTP Exchange Factor (RhoGEF) which activates Rac1 and Cdc42 (91). As adhesions mature, FAK activity becomes dispensable, but now adhesions rely on ARHGEF7 and RhoA GTPase (91). Importantly, YAP/TAZ activity can regulate FA dynamics by altering the levels of different FA components (109). Although it remains to be formally shown, these observations suggest that by having different types of YAP dynamics downstream of different adhesion types, cells can tune transcriptional events to match ECM stiffness and geometry. Critically, it has been shown that these systems that couple FA dynamics to YAP activation are often highly altered in cancer cells, emphasizing that cancer cells have evolved mechanisms such that fate determination decisions differ compared to normal cells in response to the same mechanical and geometric cues (91).

Recent modelling work has provided insight into the pathways linking mechanical cues to transcription, and how these may differ in cancer cells. A model of YAP/TAZ mechanosensing showed that YAP/TAZ increases in a switch-like manner with stiffness, and the location and plateau value of YAP/TAZ concentrations can be critically affected by the molecular state of the cell (Fig. 3c) (110). FAK is predicted to shift the location of the switch in the YAP/TAZ stiffness-response curve, whereas mDia is predicted to shift the YAP/TAZ plateau level at high stiffness. However, as discussed above, the role of FAK on YAP/TAZ signaling is complex, and coupling a model of YAP/TAZ regulation to one investigating the intricate details of FAK mechanosensing, as done in (111, 112) may provide insights toward some of these complexities.

Some mechanical effects may also underlie the behavior of multiple pathways. For instance, MRTF's are also sensitive to mechanosensing pathways, mainly through the effect of these pathways on the actin pool (113). In this way, there is some indirect overlap with YAP/TAZ signaling, but there are also direct crosstalks as discussed in (114). Further, matrix stiffness was shown to be sensed by TWIST1-G3BP2, which subsequently initiates epithelial-mesenchymal transitions (115). Matrix stiffness may also change the structure of the nucleus, in part through the coupling of the dynamics of myosin motors and lamin A, as investigated in the model in (116). This model highlights that such effects arise through general mechanisms: stable mechanosensitive gene expression may arise if a structural protein positively regulates its own gene expression while stresses inhibit the degradation of that protein. Lamin A is thus involved in mechanosensing in general and thus also affects YAP/TAZ. The typical softness of nuclei in cancer (117) may thus play a role in the altered YAP/TAZ signaling that contributes to increased malignancy (103). Recent work also showed that a direct coupling of forces through the cytoskeleton from focal adhesions to the nucleus is involved in YAP/TAZ nuclear translocation (118).

Mathematical models have also been used to predict how cell shape influences signaling dynamics. For instance, Rho GTPases activate primarily on the plasma membrane, so that shape changes will affect the effective activation rates of these GTPases as well as subsequent downstream effects (28, 29). This, for instance, implies that cells in an identical chemical state but with different shapes may react differently to chemotactic signals (Fig. 3d) (28). Moreover, since cell shape affects cytoskeletal regulators, changing shape is expected to induce feedback on shape regulation. Similarly, modeling revealed that the cAMP/PKA/B-Raf/MAPK1,2 network in neurons is controlled by cell shape (27), making cell shape a physical variable used to store biological information (119). Given the enormous heterogeneity of cellular shapes in tumors, it is thus likely that these shapes also directly contribute to the dynamics of intracellular signaling pathways and thus the heterogeneity of cell phenotypes in cancer.

## Conclusions and Outlook

Mathematical models with realistic mechanical and biochemical features have revealed underlying mechanisms and predictive insights toward how cytoskeletal components coordinate dynamically to lead to physical behaviors (migration, shape, force generation) of interest in the field of cancer biophysics. Moreover, models that directly incorporate experimentally observable or controllable features, such as dynamic adhesions, actin turnover, motors activity, and signaling, can facilitate the validation of model predictions. Further, mathematical models help provide insights toward a variety of phenomena across multiple scales, from how forces affect molecular binding rates to how tissue level stresses impact tumor progression. While many models are complex and may be computationally expensive to simulate, advances in modeling techniques, computational algorithms, and higher performance computing will enable the development of multiscale, multiphysics models that can provide an integrated picture of the various scales and features of cancer.

A key area of opportunity is the integration of models consisting of complex mechanics and biochemical signaling networks, including feedback mechanisms. Biochemical signaling networks typically have many interacting components with feedback between many pathways. Quantifying the mathematical nature of pathways that lead to cytoskeletal responses and pathways that respond to mechanotransduction can facilitate their coupling to biophysical and biomechanical models. These models can entail complex physical relationships (e.g. non-linear stiffening, viscoelasticity and plasticity) that govern discrete cytoskeletal and extracellular components (e.g. protein fibers) or the continuum representations of large quantities of these components. Novel mechanical features to incorporate that play important roles include the turnover kinetics of cytoskeletal components and the active remodeling of the cytoskeleton and ECM by molecular motors and cells. Different timescales should also be considered, from short term (minutes to hours) mechanosensing responses that lead to altered cell morphologies and cell migration directionality to long term (hours to days) mechanotransduction that leads to altered gene expression and cell fates – the mechanistic principles underlying these phenomena are not fully understood. Moreover, macroscopic tumor growth and remodeling of the ECM and metastasis may occur on even longer timescales (months to years), leading to tissue level changes of mechanics. Mathematical models aimed at understanding the interplay of mechanical processes at these vastly different time scales can help link information obtained from experiments at the molecular or cellular scale with *in vivo* or clinical observations of the long-term evolution of tumors. Coupled mechanical and systems biology models can ultimately facilitate the design of therapeutic strategies aimed toward modulating cancer phenotypes with known biophysical features, such as migratory plasticity, remodeled ECMs, and metastasis. The intersection between signaling and mechanics can provide new treatment methods against cancer, such as inhibiting or desensitizing the link between external mechanical cues (e.g. ECM stiffness) and the affected signals that drive cancer invasion and transformation (e.g. YAP/TAZ-linked pathways, integrin-mediated signals) or suppressing the pathways that lead to aggressive remodeling of the ECM. Less obvious strategies, such as targeting actin turnover to modulate cell force generation, can also be elucidated by models to guide more subtle methods toward reverting invasive phenotypes. While many drugs, such as classical microtubule-targeting chemotherapeutics or targeted inhibitors of integrins or FAK, have a clear impact on cell mechanics, little is known about the systems level effects of such drugs in dependence on the physical microenvironment of cancer. The merger of mechanical and systems biology will lead to predictive models that can help devise treatment strategies to overcome the adverse effects of mechanics on tumor progression and therapeutic resistance.

# Acknowledgements
M.M. acknowledges funding from Yale University. The content is solely the responsibility of the authors.